\documentclass[sigconf]{acmart}

\AtBeginDocument{%
  }
\setcopyright{acmlicensed}
\copyrightyear{2026}
\acmYear{2026}
\setcopyright{cc}
\setcctype{by}
\acmConference[CF '26]{Proceedings of the 23rd ACM International Conference on Computing Frontiers}{May 19--21, 2026}{Catania, Italy}
\acmBooktitle{Proceedings of the 23rd ACM International Conference on Computing Frontiers (CF '26), May 19--21, 2026, Catania, Italy}
\acmDOI{10.1145/3801487.3801824}
\acmISBN{979-8-4007-2568-5/2026/05}
\usepackage{algorithm}
\usepackage{comment}
\usepackage[noend]{algpseudocode}

\usepackage{balance}
\usepackage{tikz}
\usetikzlibrary{arrows.meta,positioning,fit,calc,shapes.geometric}

\begin{document}
%\title{HyperX: An Accurate and Energy-Efficient FPGA Accelerator for Hyperdimensional Graph Classification at the Edge}
\title{Efficient and Accurate Graph Classification with Hyperdimensional Computing on FPGA}

\author{Jebacyril Arockiaraj}
\email{arockiar@usc.edu}
\affiliation{
  \institution{University of Southern California}
  \city{Los Angeles}
  \country{USA}
}

\author{Dhruv Parikh}
\email{dhruvash@usc.edu}
\affiliation{
  \institution{University of Southern California}
  \city{Los Angeles}
  \country{USA}
}

\author{Viktor Prasanna}
\email{prasanna@usc.edu}
\affiliation{
  \institution{University of Southern California}
  \city{Los Angeles}
  \country{USA}
}

\begin{abstract}
Real-time, energy-efficient inference on edge devices is essential for graph classification across a range of applications. Hyperdimensional Computing (HDC) is a brain-inspired computing paradigm that encodes input features into low-precision, high-dimensional vectors with simple element-wise operations, making it well-suited for resource-constrained edge platforms. Recent work enhances HDC accuracy for graph classification via Nyström kernel approximations. Edge acceleration of such methods faces several challenges: (i) redundancy among (landmark) samples selected via uniform sampling, (ii) storing the Nyström projection matrix under limited on-chip memory, (iii) expensive, contention-prone codebook lookups, and (iv) load imbalance due to irregular sparsity in SpMV.

To address these challenges, we propose \textbf{HyperX}, the first end-to-end FPGA accelerator for Nyström-based HDC graph classification at the edge. HyperX integrates four key optimizations: (i) a hybrid landmark selection strategy combining uniform sampling with determinantal point processes (DPPs) to reduce redundancy while \emph{improving} accuracy; (ii) a streaming architecture for Nyström projection matrix maximizing external memory bandwidth utilization; (iii) a minimal-perfect-hash lookup engine enabling $O(1)$ key-to-index mapping; and (iv) sparsity-aware SpMV engines with static load balancing. 
Implemented on an AMD Zynq UltraScale+ (ZCU104) FPGA, \textbf{HyperX} achieves $6.85\times$ ($4.32\times$) speedup and $169\times$ ($314\times$) energy efficiency gains over optimized CPU (GPU) baselines, while \emph{improving} classification accuracy by $3.4\%$ on average across TUDataset benchmarks, a widely used standard for graph classification.

%Together, these innovations enable real-time, energy-efficient inference on resource-constrained platforms. 

\end{abstract}


%\ccsdesc[300]{Computing methodologies~Machine learning}

\maketitle

\section{Introduction}
Graphs are a natural representation for capturing relationships between entities in diverse domains such as bioinformatics~\cite{Borgwardt2005}, chemistry~\cite{molecule}, cybersecurity~\cite{Hu_Zhang_Shi_Zhou_Li_Qi_2019}, and social network analysis~\cite{10.1145/2783258.2783417}. Graphs exhibit irregular structure and variable size, making them difficult to process using conventional machine learning techniques. As a result, graph classification has emerged as a critical task and has received significant attention from the scientific community in recent years~\cite{errica2022faircomparisongraphneural}. Graph classification assigns a label to an entire graph, such as a molecule or biological structure, based on its topology and node attributes. There is a growing demand to bring such capabilities onto the edge for applications such as molecular toxicity screening, wearable biochemical sensing~\cite{rasool2025enhancing, GrapHD, NysHD}, which require real-time decisions under limited energy budgets~\cite{edge_intelligence_1, edge_intelligence_2}.

Graph Neural Networks (GNNs) are a class of models that extend deep learning to graphs through techniques such as iterative message passing and neighborhood aggregation, achieving high accuracy in tasks such as graph classification. However, they incur significant computational and energy costs, making them less suitable for resource-constrained edge devices~\cite{10.1145/3240765.3243473}.

Hyperdimensional Computing (HDC)~\cite{Kanerva2009HD} is a brain-inspired computing paradigm that represents information using low-precision vectors, often comprising thousands of dimensions. % This representation provides inherent robustness to noise. 
Unlike deep neural networks, HDC does not require complex iterative training and often generalizes well with minimal data and computation, enabling applications such as biosignal analysis and cybersecurity~\cite{static-encoding-1-programmable-hdc, static-encoding-2-seizure-det, static-encoding-3-colal-learning-secure-hdc, static-encoding-6-genie-hd, static-encoding-7-efficient-human-activity-recognition-hdc-imani, DistriHD, static-encoding-9-hdc-framework-image-descriptors-cvpr, static-encoding-11-imani-scalable-edge-hdc}. The lightweight, low complexity operations make HDC a compelling alternative to deep learning for low-power, real-time, and edge inference~\cite{10.1145/3489517.3530669, 10.1145/3495243.3558757}.  

A key step in HDC is feature encoding, which converts raw input features into hypervectors (HVs). 
Early graph-focused efforts~\cite{GrapHD} primarily captured graph topology and overlooked node attributes, limiting expressiveness. Recent work~\cite{NysHD} addresses this by enabling kernel functions in the encoding stage, allowing HDC to capture complex, nonlinear graph structures. 
During inference, given an input graph, the model performs multi-hop propagation, generates integer codes for each node, hashes them into the codebook, and updates the query graph histogram. The histogram is then compared against landmark histograms to form a kernel score, which is projected via Nyström projection matrix into an HV for final classification.

By incorporating kernel methods into HDC encoding~\cite{NysHD}, the accuracy gap of HDC with GNN-based models for graph classification is significantly reduced. Despite the advantages, applying this method at the edge raises several non-trivial challenges. First, general-purpose CPUs and GPUs are throughput-oriented and optimized for large-batch workloads, leading to poor utilization and high energy cost per query in edge inference. Second, the effectiveness of the Nyström method heavily depends on the diversity and number of landmark samples. The landmark count directly dictates the computational cost of kernel evaluations during inference and the associated memory footprint.
Third, the inherent sparsity in the propagation kernel evaluation results in redundant operations, irregular memory access, and load imbalance across compute units. Fourth, the code–to–index lookup requires accessing hop-specific codebooks; these accesses are irregular and, when executed concurrently, they contend for memory access.

FPGAs are well-suited for HDC inference as they can exploit fine-grained parallelism and implement customized data paths. 
%with banked memory access.
%These capabilities enable low-latency execution under tight resource constraints, making FPGAs a natural fit for real-time graph classification at the edge. 
In this paper, we present the first FPGA accelerator for HDC-based graph classification, delivering real-time inference for edge deployment.
%Our design explicitly tackles the unique algorithmic and architectural challenges of Nyström-based HDC.
The key contributions of this work are: 
\begin{itemize}
    \item We propose a novel sampling method that combines uniform sampling with DPP to reduce the number of redundant landmarks while preserving diversity, reducing kernel evaluation overhead and memory footprint, while \emph{improving} accuracy.
    \item Guided by our roofline analysis, we develop a streaming architecture that maximizes external memory bandwidth utilization by issuing contiguous reads matching the memory-interface width and overlapping fetch and compute.% via a FIFO.
    \item We design a minimal-perfect-hash (MPH) engine that maps computed codes (keys) to query histogram indices in $O(1)$ time, enabling parallel histogram updates. %with negligible on-chip memory overhead.
    \item We design SpMV processing elements (PEs) that exploit the sparsity in adjacency and landmark histogram matrices, reducing memory footprint and energy consumption. We employ a static load-balancing mechanism that partitions rows across PEs to reduce the load imbalance.
    \item Beyond the individual optimizations, we integrate them into \textbf{HyperX}, the first end-to-end FPGA accelerator for Nyström-based HDC graph classification at the edge, delivering real-time inference under tight memory and energy budgets.
    \item We implement our design on the AMD Zynq FPGA and evaluate it across benchmark datasets from TUDataset, a widely used standard for graph classification tasks. Our accelerator achieves $6.85\times$ ($4.32\times$) speedup and $169\times$ ($314\times$) energy efficiency over CPU (GPU) baselines, while improving classification accuracy by up to 3.4\%, on average.
\end{itemize}
\section{Preliminaries}
\label{sec:prelim}

\subsection{Background}
\label{subsec:back}
\subsubsection{Hyperdimensional Computing (HDC)}
\label{subsubsec:hdc}
\begin{comment}
HDC is a brain-inspired paradigm that represents information as high-dimensional vectors, or \emph{hypervectors} (HVs), typically with dimensionality $d \sim 10^4$~\cite{Kanerva2009HD,Kleyko2023Survey}. HVs are low-precision; in this work we assume bipolar HVs $\mathbf{h} \in \{-1,+1\}^d$. Representation learning in HDC utilizes a small set of noise-robust computations~\cite{Imani2019BRIC}: \textbf{Bundling} ($\oplus$): element-wise addition and thresholding, e.g., $\mathbf{h} = \mathrm{sign}(\mathbf{h}_1 + \mathbf{h}_2)$, to combine HVs while preserving similarity. \textbf{Binding} ($\otimes$): element-wise multiplication, $\mathbf{h} = \mathbf{h}_1 \odot \mathbf{h}_2$, to produce dissimilar HVs for encoding associations~\cite{Plate1995HRR}. \textbf{Permutation} ($\rho$): cyclic shift, $\rho^i(\mathbf{h}) = [h_{(j+i)\bmod d}]_{j=0}^{d-1}$, to encode position or sequence information~\cite{Joshi2017LanguageGeometry}.  
An HDC classifier stores \emph{class prototypes} as bundled HVs of training samples belonging to the same class. 
%often termed the model’s \emph{associative memory} (AM). 
During inference, a query HV is compared against prototypes using a similarity metric, and the class with maximum similarity is predicted. HDC thus provides 
%high noise tolerance, 
single-pass training~\cite{HernandezCano2021OnlineHD}, and hardware-friendly bit-level parallelism, making it attractive for edge-deployed learning systems~\cite{Salamat2019F5HD,Kang2022XcelHD}.
\end{comment}
Hyperdimensional Computing (HDC) is a brain-inspired paradigm that represents information as high-dimensional vectors, or \emph{hypervectors} (HVs), typically with dimensionality $d \sim 10^4$~\cite{Kanerva2009HD,Kleyko2023Survey}. HVs are low-precision; in this work we assume bipolar HVs $\mathbf{h} \in \{-1,+1\}^d$. Representation learning in HDC utilizes a small set of inexpensive, element-wise operations, and noise-robust computations well-suited for hardware acceleration~\cite{Imani2019BRIC,10.1145/3489517.3530669}:  
\begin{itemize}
    \item \textbf{Bundling} ($\oplus$): element-wise addition and thresholding, e.g., $\mathbf{h} = \mathrm{sign}(\mathbf{h}_1 + \mathbf{h}_2)$, to combine HVs while preserving similarity.  
    \item \textbf{Binding} ($\otimes$): element-wise multiplication, $\mathbf{h} = \mathbf{h}_1 \odot \mathbf{h}_2$, to produce dissimilar HVs for encoding associations~\cite{Plate1995HRR}.  
    \item \textbf{Permutation} ($\rho$): cyclic shift, $\rho^i(\mathbf{h}) = [h_{(j+i)\bmod d}]_{j=0}^{d-1}$, to encode position or sequence information~\cite{Joshi2017LanguageGeometry}.  
\end{itemize}
An HDC classifier stores \emph{class prototypes} as bundled HVs of training samples belonging to the same class. 
%often termed the model’s \emph{associative memory} (AM). 
During inference, a query HV is compared against prototypes using a similarity metric, and the class with maximum similarity is predicted. HDC thus provides 
%high noise tolerance, 
single-pass training~\cite{HernandezCano2021OnlineHD}, and hardware-friendly bit-level parallelism, making it attractive for edge-deployed learning systems~\cite{Salamat2019F5HD,Kang2022XcelHD}.

\subsubsection{Nyström Encoding for HDC}
\label{subsubsec:nystrom}
Conventional HDC encodings rely on random projections~\cite{Rahimi2007RFF,Thomas2021TheoreticalHDC, GrapHD}, which fail to capture complex data similarity. The Nyström method~\cite{Williams2000Nystrom,Kumar2012Sampling} provides a principled approach to approximate a positive semi-definite kernel $K(\cdot,\cdot)$ using a small set of $s \ll n$ \emph{landmark} samples drawn from the training data of size $n$. This enables HDC to incorporate rich kernel-based similarity functions, such as the propagation kernel, significantly improving classification accuracy for graph classification tasks~\cite{NysHD}. The propagation kernel~\cite{Neumann2016Propagation} computes similarity between graphs by iteratively propagating node features through the adjacency matrix and comparing histograms of the propagated representations. In Nyström-HDC, this process produces hop-wise histogram features that are used to compute kernel similarities to landmark graphs. Specifically, each input $x$ is represented by a kernel similarity vector $\mathbf{C}(x)\in\mathbb{R}^s$, which is projected into the hypervector space using a Nyström projection matrix $\mathbf{P}_{\mathrm{nys}}$, yielding a hypervector.

\subsection{End-to-End Nyström-HDC Inference}
\label{subsec:end-to-end-inf}
As illustrated in Algorithm~\ref{alg:end-to-end}, given a query graph $G_x=(\mathbf{A}_x,\mathbf{F}_x)$, inference uses hop-specific codebooks $\{\mathcal{B}^{(t)}\}_{t=0}^{H-1}$, landmark histogram matrices $\{\mathbf{H}^{(t)}\}_{t=0}^{H-1}$ with $\mathbf{H}^{(t)}\!\in\!\mathbb{R}^{s \times |\mathcal{B}^{(t)}|}$ (row $i$ is the hop-$t$ histogram vector of landmark $z_i$ computed during training), Locality Sensitive Hashing (LSH) parameters $\{(\mathbf{u}^{(t)}, b^{(t)})\}_{t=0}^{H-1}$ (with shared width $w>0$), the Nyström projection matrix $\mathbf{P}_{\mathrm{nys}}\!\in\!\mathbb{R}^{d\times s}$, and the class-prototype matrix $\mathbf{G}\!=\![\mathbf{g}_1;\dots;\mathbf{g}_C]\!\in\!\{-1,+1\}^{C\times d}$, to encode $G_x$ and classify it. At each hop, node features are propagated, node codes are computed in a vectorized fashion and binned (hashed) into a hop-specific histogram $\mathbf{h}^{(t)}$, which is then compared against all landmark histograms via a single matrix-vector product to yield a hop-wise similarity vector $\mathbf{v}^{(t)}\!\in\!\mathbb{R}^s$. Accumulating these over $H$ hops produces the kernel-similarity vector $\mathbf{C}\!\in\!\mathbb{R}^s$, which is embedded to an HV and matched to class prototypes. 
%Algorithm \ref{alg:end-to-end} describes the entire inference.

\begin{algorithm}
\small
\caption{End-to-End Inference for Nyström-HDC}
\label{alg:end-to-end}
\begin{algorithmic}[1]
\Require $\mathbf{A}_x\!\in\!\{0,1\}^{N\times N}$; $\mathbf{F}_x\!\in\!\mathbb{R}^{N\times f}$; hops $H$; codebooks $\{\mathcal{B}^{(t)}\}$; landmark histograms $\{\mathbf{H}^{(t)}\!\in\!\mathbb{R}^{s\times|\mathcal{B}^{(t)}|}\}$; LSH params $\{(\mathbf{u}^{(t)},b^{(t)})\}$ and width $w$; Nyström projection $\mathbf{P}_{\mathrm{nys}}\!\in\!\mathbb{R}^{d\times s}$; class prototypes $\mathbf{G}\!\in\!\{-1,+1\}^{C\times d}$
\Ensure Query HV $\mathbf{h}\!\in\!\{-1,+1\}^{d}$ and predicted class $\hat{y}$
\State $\mathbf{M}\!\gets\!\mathbf{F}_x$                                   \Comment{node features}
\State $\mathbf{C}\!\gets\!\mathbf{0}\in\mathbb{R}^{s}$                     \Comment{similarity accumulator}
\For{$t=0$ \textbf{to} $H-1$}
  \State $\mathbf{c}\!\gets\!\Big\lfloor(\mathbf{M}\mathbf{u}^{(t)}+b^{(t)}\mathbf{1}_N)/w\Big\rfloor\in\mathbb{Z}^{N}$ \Comment{LSH codes}
  \State $\mathbf{h}^{(t)}\!\gets\!\mathbf{0}\in\mathbb{Z}^{|\mathcal{B}^{(t)}|}$                                         \Comment{histogram}
  \For{$v=1$ \textbf{to} $N$}
    \State $c\!\gets\!\mathbf{c}[v]$
    \State \textbf{if} $c\!\in\!\mathcal{B}^{(t)}$ \textbf{then} $j\!\gets\!\textproc{Index}(\mathcal{B}^{(t)},c)$; $\mathbf{h}^{(t)}[j]\!\gets\!\mathbf{h}^{(t)}[j]+1$
  \EndFor
  \State $\mathbf{v}^{(t)}\!\gets\!\mathbf{H}^{(t)}\mathbf{h}^{(t)}\in\mathbb{R}^{s}$                 \Comment{hop similarity}
  \State $\mathbf{C}\!\gets\!\mathbf{C}+\mathbf{v}^{(t)}$ \Comment{accumulating into kernel similarity vector}
  \If{$t<H-1$} \State $\mathbf{M}\!\gets\!\mathbf{A}_x\mathbf{M}$ \Comment{propagate} \EndIf
\EndFor
\State $\mathbf{y}\!\gets\!\mathbf{P}_{\mathrm{nys}}\mathbf{C}\in\mathbb{R}^{d}$; \ \ $\mathbf{h}\!\gets\!\mathrm{sign}(\mathbf{y})$ \Comment{encoding}
\State $\hat{y}\!\gets\!\arg\max_{c\in\{1,\dots,C\}}\mathrm{sim}(\mathbf{h},\mathbf{g}_c)$ \Comment{similarity against class prototypes}
\State \Return $\mathbf{h}$ and $\hat{y}$
\end{algorithmic}
\end{algorithm}

\subsection{Problem Definition}
\label{subsec:probdef}
Given a labeled graph dataset $\mathcal{D}=\{(G_i,y_i)\}_{i=1}^n$ with $G_i=(\mathbf{A}_i,\mathbf{F}_i)$ and labels $y_i \in \{1,\dots,C\}$, \emph{graph classification} seeks a mapping $f:(\mathbf{A},\mathbf{F})\mapsto\{1,\dots,C\}$. At inference, a query graph $G_x=(\mathbf{A}_x,\mathbf{F}_x)$ is assigned label $\hat y=f(\mathbf{A}_x,\mathbf{F}_x)$. We specifically target \emph{single-graph, real-time} Nyström-HDC inference on resource-constrained FPGAs. For each query graph, we measure energy per graph (mJ) and end-to-end latency (ms) from input arrival to prediction. The goal is to minimize both metrics 
%on average (across a given workload) 
while operating within the constraints of on-chip memory (BRAM/URAM), compute fabric (LUTs, DSPs), and external memory bandwidth (DDR), thereby enabling real-time Nyström-HDC graph classification on edge platforms.

We summarize the computational complexity of the operations in Algorithm \ref{alg:end-to-end} in Table~\ref{tab:comp-complexity}. The table highlights that the per-graph cost is dominated by repeated propagation, codebook lookup, and landmark similarity computations across $H$ hops, along with one-time costs from Nyström projection and prototype matching. Note that $\phi_A$ and $\phi_{H^{(t)}}$ refer to the average per-row density of the sparse matrices: adjacency matrix $\mathbf{A}$ and landmark histogram matrix $\mathbf{H}^{(t)}$.

\begin{table}[ht]
\centering
\caption{Computational complexity of end-to-end Nyström-HDC inference for one query graph.}
\label{tab:comp-complexity}
\resizebox{\columnwidth}{!}{%
\begin{tabular}{@{}lccc@{}}
\toprule
\textbf{Operation} & \textbf{Expression} & \textbf{Repetition} & \textbf{Complexity} \\ \midrule
Feature Propagation & $\mathbf{M}\!\leftarrow\!\mathbf{A}\mathbf{M}$ & $H{-}1$ & $2\,\phi_A\,N^2 f$ \\
LSH Code Generation & $(\mathbf{M}\mathbf{u}^{(t)}\!+\!b^{(t)}\mathbf{1}_N)/w$ & $H$ & $2\,N f$ \\
Codebook Lookup & $c^{(t,i)} \!\in\! \mathcal{B}^{(t)}$ & $H$ & $N \log |\mathcal{B}^{(t)}|$ \\
Landmark Similarity & $\mathbf{v}^{(t)}=\mathbf{H}^{(t)}\mathbf{h}^{(t)}$ & $H$ & $2\,\phi_{H^{(t)}}\,|\mathcal{B}^{(t)}|\,s$ \\
Nyström Projection & $\mathbf{y}=\mathbf{P}_{\mathrm{nys}}\mathbf{C}$ & $1$ & $2\,s d$ \\
Prototype Matching & $\mathrm{sim}(\mathbf{h},\mathbf{g}_c)$ & $1$ & $2\,C d$ \\
Argmax & $\arg\max_c$ & $1$ & $C$ \\ \midrule
\textbf{Total} & \multicolumn{3}{l}{%
\begin{tabular}[c]{@{}l@{}}
$2(H{-}1)\phi_A N^2 f + 2H N f +\; N\!\sum_{t=0}^{H-1}\!\log|\mathcal{B}^{(t)}|$ \\
$+\; 2s\!\sum_{t=0}^{H-1}\!\phi_{H^{(t)}}|\mathcal{B}^{(t)}| +\; 2sd + 2Cd + C$ \\
\end{tabular}} \\
\bottomrule
\end{tabular}%
}
\end{table}

\section{Challenges and Key Innovation}

We summarize the key challenges and our innovations in accelerating Algorithm~\ref{alg:end-to-end} on resource-constrained edge platforms.

\noindent\textbf{Challenge \#1 --- High redundancy in landmark samples.}  
Nyström-based HDC selects landmarks via uniform sampling~\cite{NysHD}, giving every training graph an equal probability of being selected as a landmark. This often yields structurally similar landmarks, leading to redundancy. Since each query graph is compared against all landmarks during inference, this redundancy increases memory footprint and computation, scaling with $O(sd)$. \textbf{Our innovation}: A hybrid Uniform + DPP sampling strategy (offline) that maintains landmark diversity during selection (Section~\ref{sec:dpp}, \ref{results:dpp}).

\noindent\textbf{Challenge \#2 --- Large projection matrix size.}  
The Nyström projection matrix dominates model parameters and exceeds on-chip memory capacity on edge FPGAs (e.g., ZCU104 provides $\sim$4.5\,MB versus 7--16\,MB required. Consequently, the matrix must be streamed from off-chip DDR during inference with limited external memory bandwidth. \textbf{Our innovation}: Guided by our roofline analysis (Section~\ref{sec: roofline analysis}), we design a streaming architecture that maximizes external memory bandwidth (Section~\ref{sec: NEE streaming}).

\noindent\textbf{Challenge \#3 --- Expensive codebook lookup.}  
For every node and hop, the code must be mapped through a hop-specific codebook to locate its histogram bin. Naïve implementations require repeated dictionary searches, which scale poorly and introduce conflicts when many nodes are processed in parallel. \textbf{Our innovation}: We design a novel lightweight MPH-based lookup engine that delivers constant-time ($O(1)$) code-to-index mapping, by banking both hash tables and codebook, minimizing conflicts (Section~\ref{subsec:mphe}).

\noindent\textbf{Challenge \#4 --- Severe load imbalance in SpMV.}  
Sparse matrix operations such as feature propagation (via $\mathbf{A}$) and landmark histogram similarity computation (via $\mathbf{H}^{(t)}$) can become bottlenecks despite their low computational cost. Irregular row sparsity often leaves some PEs idle while others stall on dense rows, leading to severe under-utilization of compute resources. \textbf{Our innovation}: We employ a static load balancing strategy based on precomputed schedule tables (Section~\ref{subsec:static-lb}), eliminating under-utilization.% and sustaining high PE throughput.

\section{Optimizations}

\subsection{Landmark Reduction via DPP Sampling}\label{sec:dpp}

%\textcolor{red}{[Comment: Describe how Determinantal Point Processes (DPPs) are used to reduce the number of landmarks while preserving diversity. 
%Explain why fewer landmarks reduce both memory and compute costs, and how DPP sampling improves representativeness compared to random or uniform sampling. 
%Emphasize the co-design angle: this algorithmic optimization reduces footprint and latency on FPGA while also improving accuracy.]}

An important step in Nyström-based HDC encoding~\cite{NysHD} is the selection of landmark graphs~\cite{fastdpp}. These landmarks serve as the representative subset used to construct the Nyström projection matrix, and during inference, every query graph must be compared against the landmarks. The diversity of this landmark set, therefore, directly affect both the computational cost and the classification accuracy. Baseline approaches often rely on random uniform sampling~\cite{Williams2000Nystrom}, which does not guarantee diversity. As a result, graphs with a similar structure and node features may repeatedly be chosen as landmarks, increasing the projection cost while reducing accuracy due to bias toward certain structures. To address this, we adopt a diversity-aware selection method based on Determinantal Point Processes (DPPs)~\cite{Kulesza_2012}. DPPs favor subsets with high pairwise dissimilarity, thereby improving diverse landmark coverage. 
%In our approach, the DPP similarity kernel is built using the graph propagation kernel~\cite{Neumann2016Propagation}. 

Unlike prior works~\cite{fastdpp}, which target Euclidean data, where kernel evaluations reduce to dot products or distance computations, graph similarity requires propagation kernel~\cite{Neumann2016Propagation} evaluations that are substantially more expensive. Consequently, constructing the DPP kernel requires all pairwise kernel evaluation, and the DPP sampling further incurs eigen decomposition cost ($O(s^3)$). Thus, applying DPP directly to all training graphs is impractical. Therefore, we adopt a hybrid Uniform+DPP strategy that first reduces the candidate pool to a maximum of 300 graphs, or 2\% of the training set, as in~\cite{NysHD} with uniform sampling, and then applies DPP.% to further reduce the number of landmarks while preserving diversity.
%, making DPP-based landmark selection practical for Nyström-HDC on graph datasets. 

\begin{algorithm}
\small
\caption{Hybrid Landmark Selection with DPP}
\label{alg:dpp-landmark}
\begin{algorithmic}[1]
\Require Graph set $\mathcal{G}$, target landmark count $s$
\Ensure Landmark set $\mathcal{L}$
\State Draw candidate pool $\mathcal{C} \subset \mathcal{G}$ using uniform sampling
\State Compute similarity kernel $K$ over $\mathcal{C}$ using the propagation kernel
\State Apply DPP sampling on $K$ to select $s$ diverse landmarks
\State \Return $\mathcal{L}$
\end{algorithmic}
\end{algorithm}

%\textcolor{red}{As illustrated in Algorithm~\ref{alg:dpp-landmark}: we first reduce the candidate pool by uniform sampling, using a size equal to the maximum of 300 graphs or two percent of the training set, as in~\cite{NysHD}, and then apply DPP to select the final landmark set.} 
This strategy retains the diversity benefits of DPP while keeping the landmark selection cost during training manageable. The impact is twofold. First, fewer redundant landmarks reduce the dimensionality $s$ of the Nyström projection matrix $\mathbf{P}_{\mathrm{nys}} \in \mathbb{R}^{d \times s}$, which directly lowers both computation and memory usage. Second, the improved diversity of landmarks enhances classification accuracy by reducing the overrepresentation of similar graph patterns. These benefits make DPP-based landmark selection a practical optimization for deploying Nyström-HDC~\cite{NysHD} on edge FPGAs.

\subsection{Static Load Balancing}
\label{subsec:static-lb}
Irregular sparsity in adjacency matrices ($\mathbf{A}$) and landmark histograms ($\mathbf{H}^{(t)}$) can cause severe PE underutilization. To avoid runtime scheduling overhead while maintaining balanced utilization, we design a lightweight \emph{static iteration-wise load balancer} that maps rows to PEs offline, ensuring all PEs remain evenly loaded.

\emph{Schedule table.} Given $N$ rows and $P$ PEs, computation proceeds in $N/P$ iterations. In each iteration, every PE processes exactly one row, with assignments specified by a precomputed $N/P \times P$ \emph{schedule table}. Each row of the table corresponds to one iteration, and each column to a PE; entry $(i,j)$ gives the row index assigned to PE~$j$ in iteration~$i$. The table is banked along columns, allowing PEs to fetch their assigned rows concurrently without conflict. At runtime, PE~$j$ fetches its row index, reads $\texttt{row\_ptr}$ and $\texttt{col\_idx/val}$ from Compressed Sparse Row (CSR) banks, and accumulates results directly into $\texttt{out[row\_id]}$.

\emph{Offline construction.} The table is constructed once per sparse operand using a simple nonzero-count grouping procedure:
(1) Compute $\texttt{nnz}[r]$ number of non-zero elements in each row; (2) Bucket rows into lists keyed by $\texttt{nnz}$; (3) Traverse buckets in increasing order of $\texttt{nnz}$, greedily allocating $P$ rows at a time to form one iteration. If a bucket contains fewer than $P$ rows, allocation continues with the next bucket until $P$ rows are obtained. This procedure groups rows with similar sparsity into the same iteration, yielding balanced PE utilization at each cycle. It runs in $O(N)$ time, preserves the CSR layout, and produces a compact schedule table stored in BRAM.

\emph{Execution.} At runtime, the controller issues iterations sequentially. Each PE reads its row index from the schedule table, performs CSR expansion, and writes directly to $\texttt{out[row\_id]}$, with no dynamic scheduling or reassignment required.

\section{Accelerator Design}
\label{sec:method}

\subsection{HyperX Overview}
\label{subsec:ovw}

\begin{figure}
  \centering
  \includegraphics[width=\columnwidth]{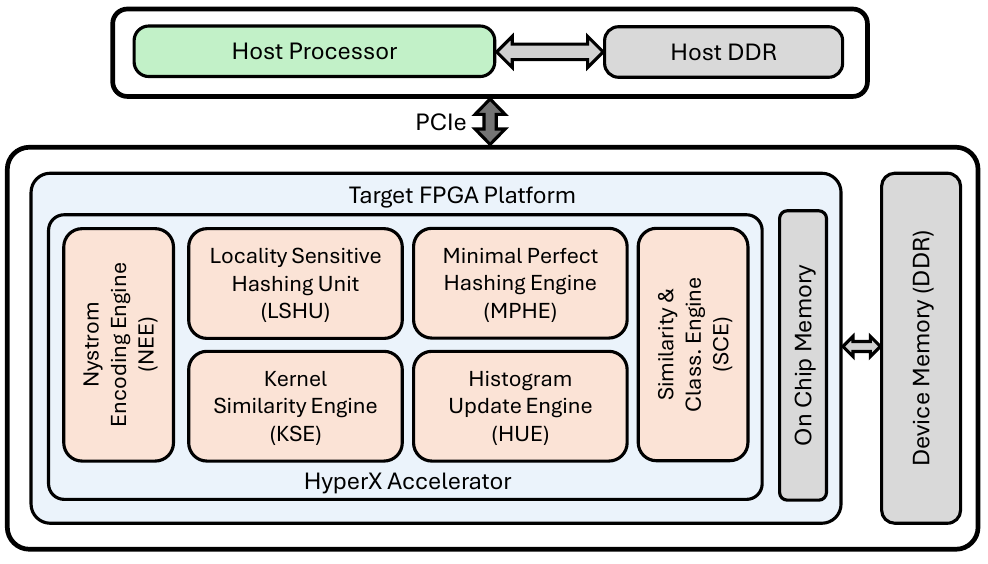}
  \caption{Overview of the HyperX FPGA accelerator.}
  \label{fig:HyperX-overview}
\end{figure}

Figure~\ref{fig:HyperX-overview} shows the overview of \textbf{HyperX}, our FPGA accelerator for Nyström-HDC graph classification. \textbf{HyperX} is composed of six compute engines: (i) the Locality Sensitive Hashing Unit (LSHU), which generates integer codes for nodes at each hop; (ii) the Minimal Perfect Hashing Engine (MPHE), which maps integer codes to valid histogram indices in $O(1)$ time using a MPH; (iii) the Histogram Update Engine (HUE), which accumulates counts into hop-specific histograms using the indices from MPHE; (iv) the Kernel Similarity Engine (KSE), which computes landmark similarities by multiplying the query histogram with buffered landmark histograms; (v) the Nyström Encoding Engine (NEE), which projects the kernel vector into the hyperdimensional space; and (vi) the Similarity \& Classification Engine (SCE), which computes final class scores by comparing the encoded hypervector against class prototypes, followed by argmax selection of the label.
%On-chip BRAM stores hash level tables, landmark histograms, and intermediate results. The Nyström projection matrix, due to its large size, is streamed from external DDR memory.

%A lightweight controller orchestrates hop iteration between engines.

\subsection{Architectural Components}

\subsubsection{\textbf{Locality Sensitive Hashing Unit (LSHU)}}
\label{subsubsec:lshu}

The LSHU is responsible for generating hop-specific codes. As depicted in Algorithm~\ref{alg:end-to-end}, Nystrom-HDC inference first computes the locality sensitive hash $\Big\lfloor \frac{\mathbf{M} \mathbf{u} \;+\; b \mathbf{1}_N}{w} \Big\rfloor$, where $\mathbf{M} = \mathbf{F}_x$ initially, and then propagates $\mathbf{M} \leftarrow \mathbf{A} \mathbf{M}$, to generate code vector $\mathbf{c}$. This requires storing intermediate feature matrices of size $O(Nf)$. In our implementation, we restructure this computation into a sequence of matrix-vector products, computed via LSHU. Specifically, we first compute the projection $\mathbf{c} = \mathbf{F}_x \mathbf{u}$. Next, we iterate over the hop count to compute $\mathbf{c} \leftarrow \mathbf{A} \mathbf{c}$. This computes the required $\mathbf{M}^{(k)}\mathbf{u}^{(k)}$ as $\mathbf{A}^{k} \mathbf{F}_x \mathbf{u}^{(k)}$. While both methods require access to $\mathbf{F}_x$, the baseline requires storing the full feature matrix $\mathbf{M}^{(k)} \in \mathbb{R}^{N \times f}$ as opposed to the intermediate vectors of length $N$ required by our restructuring, significantly reducing the on-chip memory resources required during code generation. In terms of complexity, the baseline requires $HNf + (H-1)f\,\text{nnz}(\mathbf{A}_x)$ operations, whereas our restructuring reduces this to $HNf + \tfrac{H(H-1)}{2}\,\text{nnz}(\mathbf{A}_x)$. This offers a clear advantage when $f > H/2$, which typically holds since hop counts are small ($H \leq 10$) while feature dimensions are much larger (e.g., $f \approx 50$).

\begin{figure*}
  \centering
  \includegraphics[width=\linewidth]{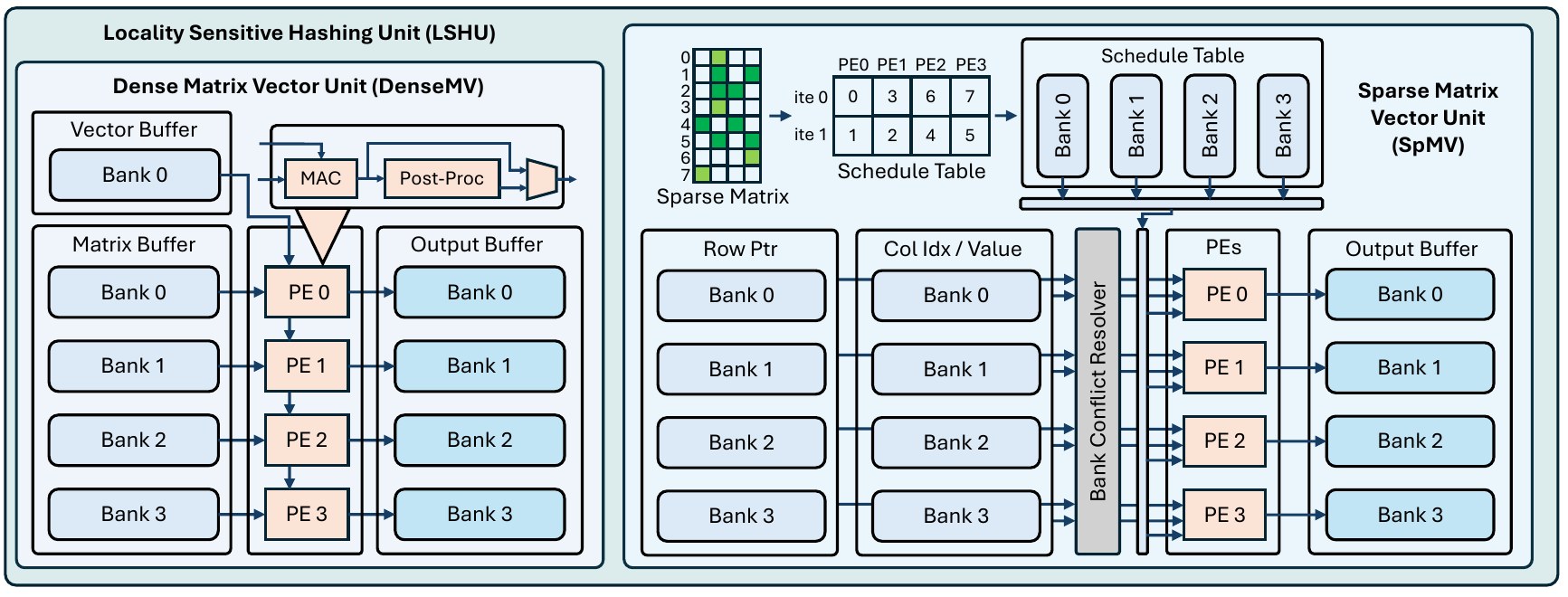}
  \caption{Locality Sensitive Hashing Unit (LSHU). It integrates a DenseMV unit (for $\mathbf{F}\mathbf{u}$) and a SpMV unit (for $\mathbf{A}\mathbf{c}$), with schedule tables, CSR arrays, and output buffers banked across BRAMs; a bank conflict resolver manages access overlaps, enabling concurrent output computation.}
  \label{fig:lshu}
\end{figure*}

Real-world graphs are sparse~\cite{GrapHD}, and the LSHU exploits this by storing $\mathbf{A}$ in CSR format and skipping zero entries, thereby reducing both the compute and memory bandwidth demand — critical under edge constraints. However, irregular sparsity causes load imbalance across PEs. To mitigate this, we employ \emph{static schedule tables} (Section~\ref{subsec:static-lb}), which pre-assign rows to PEs based on nonzero counts, ensuring balanced utilization without runtime scheduling overhead. Figure~\ref{fig:lshu} summarizes the architecture. The DenseMV unit performs feature projection, while the SpMV unit handles hop-wise propagation. For SpMV, schedule tables are banked and accessed concurrently by all PEs, enabling each PE to fetch its assigned rows in parallel. CSR arrays (\texttt{row\_ptr}, \texttt{col\_idx}, \texttt{val}) and outputs are also banked across BRAMs to reduce contention, with conflicts managed by a bank conflict resolver. Each PE includes a dedicated MAC unit with a private accumulator and writes results directly to its assigned output index, enabling parallel and energy-efficient sparse computation.

\subsubsection{\textbf{Minimal Perfect Hashing Engine (MPHE)}}
\label{subsec:mphe}
We adopt a MPH scheme~\cite{BBHash} to map codes (keys) in the codebook to indices $\{0,\dots,|\mathcal{B}^{(t)}|\!-\!1\}$ with $O(1)$ query time during inference. The scheme constructs the MPH as a cascade of \emph{levels} $A_0,A_1,\dots$, where each bit-array $A_d$ is sized to balance memory against collision probability. During construction, keys hashing uniquely at level $d$ set a $1$ at position $h_d(\text{key})$ and are removed, while colliding keys advance to the next level $d{+}1$. The final structure concatenates all bit-arrays with a lightweight \emph{rank} vector. At query time, a key traverses the levels until it encounters a $1$, and its index is computed as the cumulative number of 1s preceding that position. This yields a compact ($\approx\!3$ bits/key) structure supporting constant-time queries.
The MPHE performs inference-time dictionary lookup from code $\rightarrow$ histogram index. Upstream, the LSHU produces a chunk of integer codes, which are enqueued into a \emph{lookup queue}. Each lookup then proceeds as follows:
\begin{enumerate}
    \item \emph{Hash generation.} A \emph{Hash Function Engine} computes two 64-bit hashes for the input code using a seeded integer hash function \cite{integerhash}. The level-$d$ probe index is $i_d \gets h_d(\text{code}) \bmod |A_d|$, where $|A_d|$ is the bit capacity of level table $A_d$. 
    \item \emph{Level table probe.} Level tables $\{A_d\}$ are stored on-chip as banks of 64-bit words; the bit at $i_d$ is checked within its word. If it is 0, the code advances to the next level by generating the next hash in the sequence via a xorshift-based rehash generator \cite{sequencehashgen}. 
    \item \emph{Rank to MPH index.} On a 1-bit hit at $(d,i_d)$, the engine retrieves the global index directly from the rank vector. Each entry in the rank vector stores the cumulative number of 1s up to the start of a word, aggregated across all levels. The final index is obtained by adding this value to the popcount of bits within the current word up to (and including) position $i_d$, and subtracting one for zero-based indexing. 
    \item \emph{Codebook verification.} The MPH index addresses a compact \emph{codebook store} that holds pairs \texttt{(code,hist\_idx)}. The queried code is compared against the stored code. If they match, the corresponding \texttt{hist\_idx} is emitted as valid. If they differ, the lookup terminates, as a hit with a mismatch implies the queried code is not in the original codebook. If all levels are probed with no hit, the code is deemed absent. No histogram update occurs for a queried code absent from the codebook.
\end{enumerate}

%\noindent \textbf{Microarchitecture.} 

\begin{figure}
  \centering
  \includegraphics[width=\columnwidth]{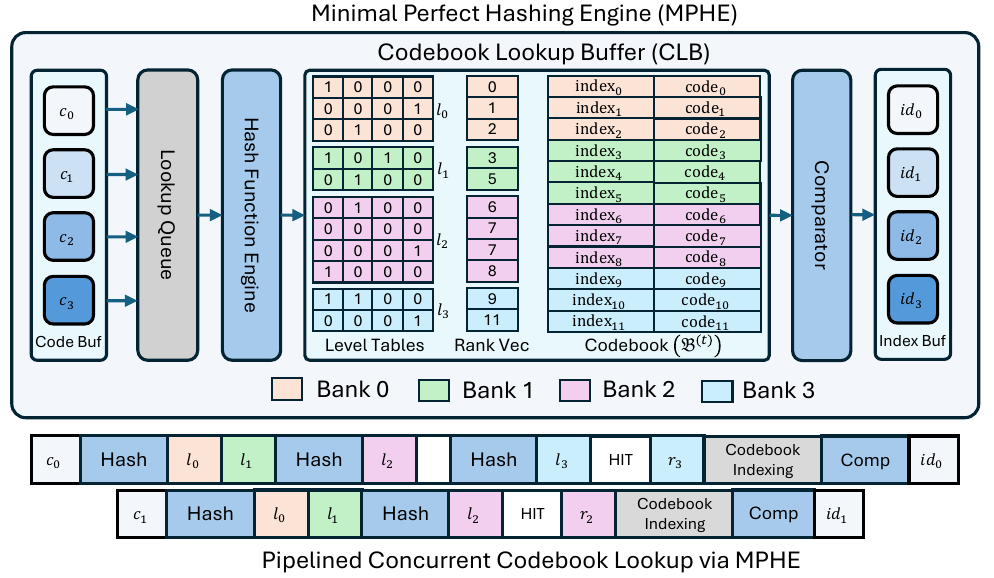}
  \caption{Minimal Perfect Hashing Engine (MPHE). Codes from LSHU PEs are hashed and probed via banked codebook lookup buffer (CLB). Multiple lookups proceed in a pipelined manner (bottom), with $O(1)$ query time.}
  \label{fig:mphe}
\end{figure}

Figure~\ref{fig:mphe} illustrates the MPHE datapath: (a) a \emph{Lookup Queue} collects codes from LSHU PEs; (b) a \emph{Hash Function Engine} generates 64-bit hashes and level indices; (c) \emph{Level Tables} $A_d$ and \emph{Rank Vectors}, each banked across BRAMs, are accessed in parallel; (d) an \emph{Index Buffer} carries the computed MPH index; and (e) a banked \emph{Codebook} returns the \texttt{(code,hist\_idx)} pair to a \emph{Compare Unit} that verifies code presence. Level tables and rank vectors are placed in independent banks to support pipelined parallel lookups (Figure \ref{fig:mphe}), while contention in codebook access is reduced through MPH hashing, pipelining, and banking. The pipeline roughly issues one index lookup per cycle, with $O(1)$ cost bounded by the small number of level tables per codebook. 

\subsubsection{\textbf{Histogram Update Engine (HUE)}}
\label{subsec:hue}

The HUE maintains hop-specific query histogram vector $\mathbf{h}^{(t)}$, incrementing it based on the indices returned by MPHE. To maximize update parallelism, each PE in HUE keeps a private copy of the histogram, allowing concurrent increments without contention. Once all codes in the current hop are processed, local histograms are merged into the global hop histogram through a reduction step. This two-stage strategy avoids write conflicts. % while still providing correct cumulative counts.
Each histogram is stored in on-chip BRAM, with indexing controlled by MPHE outputs. The final merged histogram is forwarded to KSE.% for landmark similarity computation.

\subsubsection{\textbf{Kernel Similarity Engine (KSE)}}
\label{subsec:kse}

The KSE computes hop-level landmark similarities by multiplying the query histogram $\mathbf{h}^{(t)}$ with the precomputed landmark histograms $\mathbf{H}^{(t)}$, producing $\mathbf{v}^{(t)} = \mathbf{H}^{(t)} \mathbf{h}^{(t)}$. This operation is realized using a sparse matrix–vector (SpMV) unit, as $\mathbf{H}^{(t)}$ is stored in CSR format and exhibits significant sparsity. Rows of $\mathbf{H}^{(t)}$ are distributed across PEs, and static load balancing (Section \ref{subsec:static-lb}) ensures uniform work assignment. The resulting vector $\mathbf{v}^{(t)}$ is accumulated into the global kernel similarity buffer $\mathbf{C}$ across all hops.

\subsubsection{\textbf{Nyström Encoding Engine (NEE)}}

The NEE generates the final hypervector through projection $h = \mathrm{sign}(\mathbf{P}_{\mathrm{nys}} \mathbf{C})$, where $\mathbf{P}_{\mathrm{nys}}$ is the Nyström projection matrix, $\mathbf{C}$ is the similarity vector from the KSE compute engine. Storing $\mathbf{P}_{\text{nys}}$ on-chip is infeasible for edge devices with limited on-chip memory; hence, $\mathbf{P}_{\mathrm{nys}}$ is streamed from external DDR during inference. Profiling shows that the NEE dominates inference time (>90\%), making it a critical engine, and we analyze it using the roofline model~\cite{roofline}. 
%to identify whether performance is constrained by compute or by memory bandwidth.

\textbf{Roofline Analysis:}\label{sec: roofline analysis}
%To understand the performance limits of the NEE, we analyze it using the Roofline model. 
%The Roofline analysis relates three quantities: arithmetic intensity (AI), machine balance (B), and peak performance. Arithmetic intensity is defined as the ratio of operations performed to bytes transferred from memory. Machine balance is the ratio of the device’s peak performance to its peak memory bandwidth. A kernel is considered memory-bound if its AI is lower than the machine balance, meaning performance is limited by how fast data can be moved from memory rather than by raw compute capacity.
For the NEE projection $\mathbf{P}_{\mathrm{nys}} \mathbf{C}$, while $\mathbf{C}$ is buffered on-chip, each multiply-accumulate reads one element of $\mathbf{P}_{\mathrm{nys}}$ matrix and performs two floating-point operations. With 32-bit elements (4 bytes), this gives ~2 ops per 4 bytes, which translates to an arithmetic intensity (AI) of 0.5 operations per byte (ops/byte). To illustrate machine balance, consider a design point with 32 MAC lanes clocked at 300 MHz, yielding a peak compute performance of 19.2 GOPS. The ZCU104’s DDR4 interface provides a theoretical bandwidth of 19.2 GB/s~\cite{10.1145/3431920.3439284}; even at 90\% efficiency, sustained bandwidth is ~17.3 GB/s. This corresponds to a machine balance of about 1.11 ops/byte. Since AI is less than machine balance, the kernel falls into the memory-bound region of the roofline model. This implies that performance gains primarily come from improving data movement rather than simply adding more MAC lanes.

\begin{figure}
  \centering
  \includegraphics[width=\columnwidth]{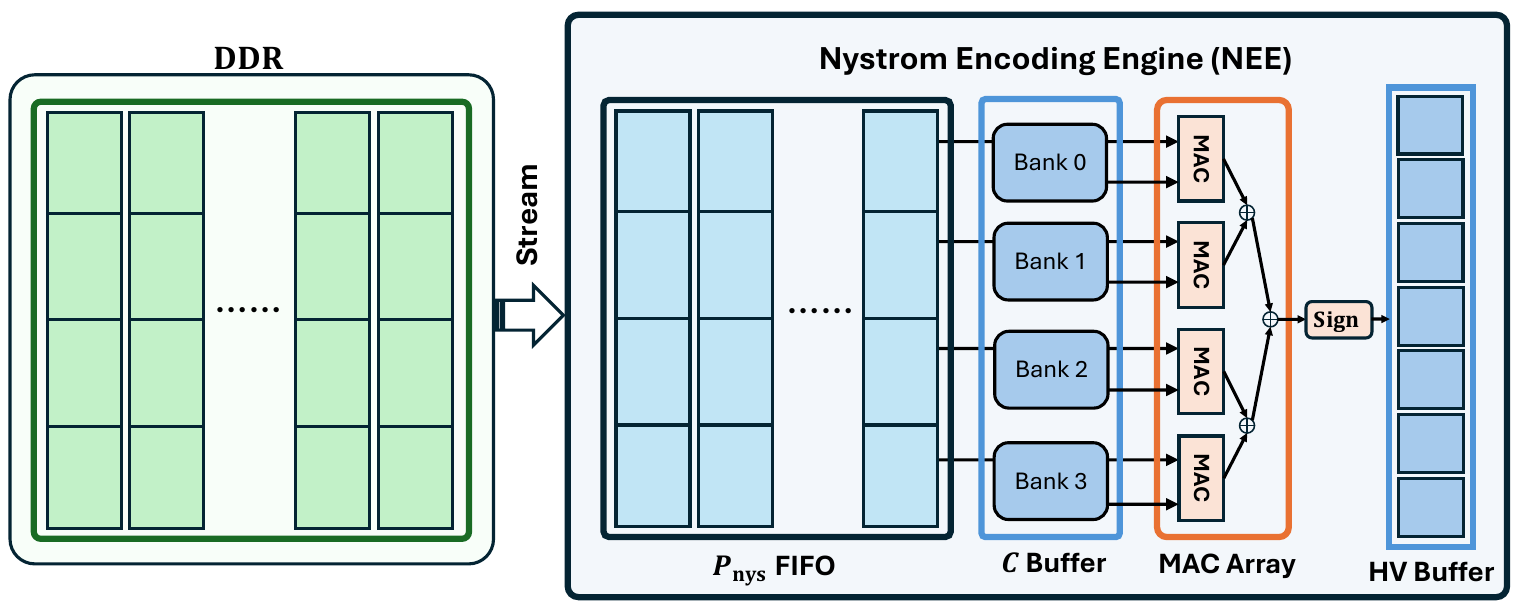}
  \caption{Streaming Nyström Encoding Engine (NEE) that overlaps DDR reads through a FIFO into parallel MAC lanes, multiplies with banked $C$ vector, and outputs thresholded bipolar hypervectors into HV buffer.}
  \label{fig:nee}
\end{figure}

\begin{figure*}
  \centering
  \includegraphics[width=0.9\linewidth]{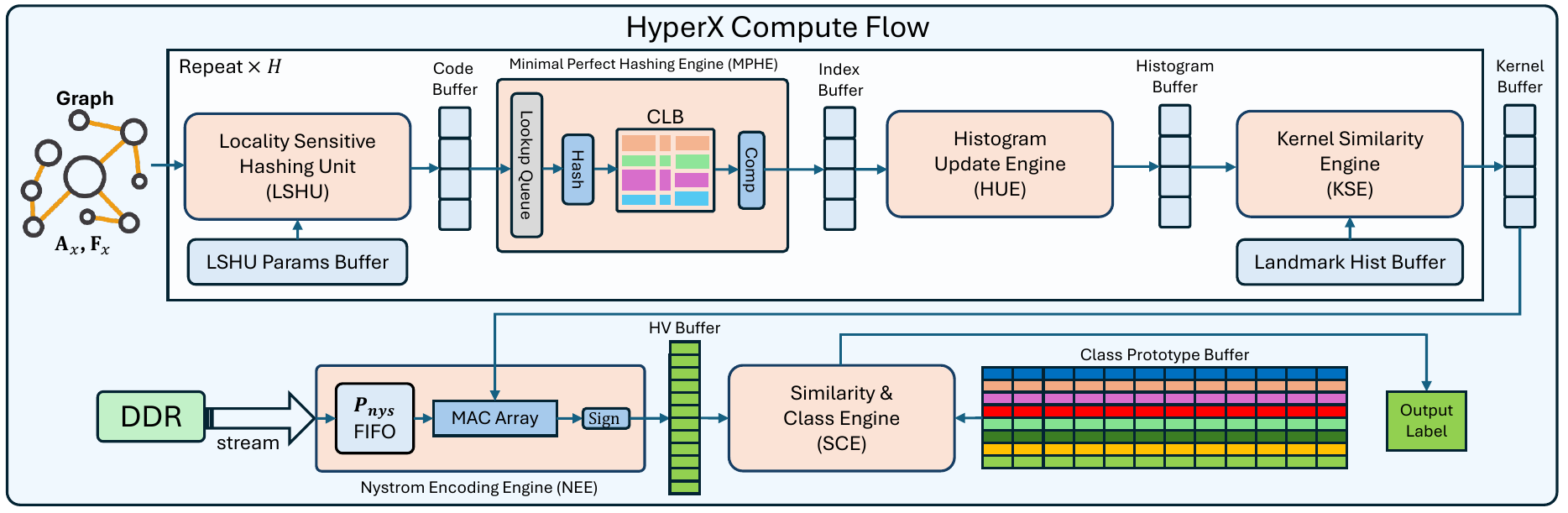}
  \caption{HyperX compute flow across hops: input graph $\{A_x,F_x\}$ is processed by compute engines to produce class label.}
  %LSHU, MPHE, HUE, KSE, NEE, and SCE to produce the output class label.}
  \label{fig:compute-flow}
\end{figure*}

\textbf{Streaming Architecture:}\label{sec: NEE streaming} Given that the NEE projection is memory-bound, this work adopts a streaming (dataflow)~\cite{streaming_arch} architecture in which compute stages are connected via on-chip FIFOs to overlap data movement with execution. 
%we maximize sustained memory bandwidth using the streaming architecture in Figure~\ref{fig:nee}.
To maximize external memory bandwidth utilization, data transfers are aligned to the FPGA platform’s native memory-port width, $y$ bits. Accordingly, we organize $\mathbf{P}_{\mathrm{nys}}$ into $y$-bit words and issue contiguous $y$-bit burst transfers. This maximizes effective bandwidth utilization, avoiding inefficiency from irregular or undersized accesses. We also enable multiple outstanding reads to keep the memory interface saturated despite variable off-chip DRAM latency. Incoming $y$-bit words are buffered in a deep on-chip FIFO, decoupling bursty DRAM traffic from the compute stage. 
%Without this buffering, memory-interface stalls would propagate into the MAC pipeline, reducing throughput. 
Each $y$-bit word is unpacked into $y$/$x$ operands, where $x$ is the precision of the operand in bits, and is routed to $y$/$x$ MAC lanes, fully utilizing the memory interface width. As illustrated in Figure~\ref{fig:nee}, each MAC lane performs a dot product between its assigned elements of $\mathbf{P}_{\mathrm{nys}}$ and the corresponding entries of the similarity vector $\mathbf{C}$, which is buffered on chip with cyclic partitioning to enable conflict-free parallel access. After accumulation, results are bipolarized to $\{-1,+1\}$.

\subsubsection{\textbf{Similarity \& Classification Engine (SCE)}}
\label{subsec:sce}

The SCE performs the final stage of inference by comparing the encoded hypervector $\mathbf{h}$ against the stored class prototypes $\{\mathbf{g}_c\}_{c=1}^C$. This is implemented as a dense matrix–vector multiplication $\mathbf{s} = \mathbf{G} \mathbf{h}$, where $\mathbf{G} \in \{-1,+1\}^{C \times d}$. The computation is parallelized across PEs, each responsible for a block of rows in $G$. The output scores $\mathbf{s}$ are written into an on-chip buffer, followed by a small argmax unit that identifies the predicted class $\hat{y} = \arg\max_c s_c$.

\subsection{Compute Flow}

Figure~\ref{fig:compute-flow} illustrates the end-to-end execution flow of \textbf{HyperX}, tying together the specialized compute engines described earlier. The process begins with an input graph, represented by adjacency matrix $\mathbf{A}_x$ and feature matrix $\mathbf{F}_x$. Inference proceeds hop-by-hop, with each hop producing a histogram vector that contributes to the global kernel similarity vector. At hop $h$, the \emph{LSHU} computes code vectors. First, a dense MV unit projects features as $\mathbf{c} \gets \mathbf{F}_x \mathbf{u}^{(h)}$. This is followed by sparse propagation $\mathbf{c} \gets \mathbf{A}_x \mathbf{c}$, performed iteratively. Exploiting CSR format and static load balancing, each PE processes rows of $\mathbf{A}_x$ with comparable nonzero counts. Floor operation then produces integer codes, which are enqueued for dictionary lookup. 

Codes are passed to the \emph{MPHE}, which maps each code to a histogram index via pipelined, $O(1)$ lookups. Level tables, rank vectors, and codebook stores are banked in BRAM, enabling concurrent access across PEs. Verified histogram indices are forwarded to the \emph{HUE}, where PEs increment local histograms in parallel before merging into the hop-global histogram buffer. Once a hop histogram is complete, the \emph{KSE} multiplies it with pre-stored landmark histograms. Implemented as a SpMV, this step accumulates partial results into the global kernel similarity vector $\mathbf{C}$. After all hops finish, the \emph{NEE} consumes $\mathbf{C}$ from the kernel buffer. The projection matrix $\mathbf{P}_{\text{nys}}$ is streamed from DDR through a FIFO into the MAC array, where they are multiplied with the corresponding elements of $\mathbf{C}$ to generate the output hypervector. The $\mathrm{sign}()$ function is fused directly into the MAC array, producing a bipolarized HV on the fly. 
%This streaming design integrates projection and quantization in one pass, reducing bandwidth demand by eliminating the need to store intermediate high-precision values.
Finally, the \emph{SCE} computes class scores by multiplying the encoded hypervector with class prototypes, followed by an argmax to select the label.

\section{Experimental Evaluation}

\subsection{Implementation Details}
We implement the proposed accelerator on an AMD Zynq UltraScale+ ZCU104 FPGA~\cite{zcu104}, representative of an edge-class FPGA. 
%The design targets real-time inference with batch size~1, consistent with edge deployment scenarios where query graphs arrive sequentially.
The accelerator is designed using AMD Vitis High Level Synthesis (HLS) and synthesized using AMD Vitis Unified IDE v2024.2~\cite{AMD_Vitis_Unified_IDE_2024_2}. We instantiate 4 PEs in the LSHU, KSE, and HUE compute engines of the accelerator to conserve on-chip resources (LUT/DSP/BRAM) for the NEE compute engine, which dominates area and latency. Empirically, configurations with more than 4 PEs result in marginal speedup while increasing resource usage, so 4 PEs remain the best end-to-end trade-off on ZCU104. For NEE, $\mathbf{P}_{\mathrm{nys}}$ is streamed through a 512-bit AXI transfer width. This width is selected because Vitis HLS automatically widens burst ports up to 512 bits~\cite{AMD_Vitis_HLS_UG1399} and AXI SmartConnect inserts width converters to match the platform’s native memory ports~\cite{AMD_SmartConnect_PG247}. For a 32-bit precision, each transfer packs \(16\) FP32 values and drives \(16\) parallel MAC lanes. We employ a 512-entry, 16-FP32-wide stream FIFO to overlap data fetch and compute. Table~\ref{tab:util} reports the resource utilization obtained through the Vitis Unified IDE, with an achieved frequency of \(300\,\text{MHz}\).

\begin{table}
  \centering
  \caption{Resource Utilization of the Proposed Design}
  \label{tab:util}
  \begin{tabular}{lrrr}
    \toprule
    \textbf{Resource} & \textbf{Used} & \textbf{Available} & \textbf{Utilization} \\
    \midrule
    LUT          & 71,900  & 230,400 & 31\% \\
    FF           & 87,800  & 460,800 & 19\% \\
    BRAM (18K)   & 329     & 624     & 52\% \\
    DSP          & 156     & 1,728   & 9\%  \\
    URAM         & 0       & 96      & 0\%  \\
    \bottomrule
  \end{tabular}
\end{table}

\subsection{Datasets}

We evaluate on eight benchmark datasets from the TUDataset \cite{tudatasetcollectionbenchmarkdatasets}, a widely used collection for graph classification, consistent with prior work on HDC-based graph classification~\cite{GrapHD, NysHD}. These datasets span diverse domains (e.g., bioinformatics, chemistry, and drug discovery) and vary in graph sizes and structures, providing a representative set for evaluating \textbf{HyperX}. Importantly, graph classification benchmarks consist of many independent graphs whose sizes are inherently bounded by the application domain; large single-graph workloads (e.g., citation networks) are characteristic of node classification and are not representative of this task. Table~\ref{tab:datasets} summarizes the dataset statistics.

\begin{table}[ht]
  \centering
  \caption{Summary of Graph Classification Datasets}
  \label{tab:datasets}
  \resizebox{\columnwidth}{!}{%
  \begin{tabular}{lrrrrrr}
    \toprule
    \textbf{Task} & \textbf{\#Train} & \textbf{\#Test} & \textbf{Avg. Nodes} & \textbf{Avg. Edges} & \textbf{Description} \\
    \midrule
    ENZYMES~\cite{Borgwardt2005}      & 480  & 120  & 33  & 62  & Protein graphs \\
    NCI1~\cite{molecule}              & 3288 & 822  & 30  & 32  & Chemical compounds \\
    D\&D~\cite{Dobson2003}            & 943  & 235  & 284 & 716 & Protein structures \\
    BZR~\cite{Sutherland2003}         & 324  & 81   & 36  & 38  & Drug activity graphs \\
    MUTAG~\cite{Debnath1991}          & 150  & 38   & 18  & 20  & Mutagenicity prediction \\
    COX2~\cite{Sutherland2003}        & 373  & 94   & 41  & 43  & Drug activity graphs \\
    NCI109~\cite{molecule}            & 3301 & 826  & 30  & 32  & Chemical compounds \\
    Mutagenicity~\cite{Riesen2008}    & 3469 & 868  & 30  & 31  & Mutagenicity prediction \\
    \bottomrule
  \end{tabular}}
\end{table}

\subsection{Baseline Platforms}
We evaluate our FPGA accelerator against CPU and GPU baselines: the AMD Ryzen 5 5625U and the NVIDIA RTX A4000. All baselines are implemented in \texttt{PyTorch} (v2.4.1) with Python~3.8 and CUDA~12.1. We select the Ryzen processor as representative of CPUs deployed in laptops and compact systems, while the RTX A4000 is a workstation-class GPU commonly used as a baseline in prior HDC works~\cite{hdreason, hdnn-pim, paap-hd, survey-edge}. Both baselines offer substantially higher peak compute and memory resources than our target edge FPGA, providing a conservative comparison. For fairness, GPU latency is measured after all model parameters are transferred to device memory. This favors the GPU, which benefits from high-bandwidth GDDR6 memory, whereas on the FPGA, the Nyström projection matrix—the dominant model parameter—must be streamed from DDR during inference. Despite this, \textbf{HyperX} consistently achieves lower latency. While workstation-class GPUs naturally draw more power, we report energy efficiency (mJ/graph) to ensure fairness.

\begin{table}
  \centering
  \caption{Specifications of Baseline Platforms}
  \label{tab:baseline-platforms}
  \resizebox{\columnwidth}{!}{%
  \begin{tabular}{lccc}
    \toprule
    & \textbf{CPU} & \textbf{GPU} & \textbf{FPGA (Ours)} \\
    \midrule
    Platform & AMD Ryzen 5 5625U & NVIDIA RTX A4000 & AMD ZCU104 \\
    %Technology & TSMC 7nm & Samsung 8nm & TSMC 16nm \\
    Frequency & 2.3 GHz (base) & 1560 MHz & 300 MHz \\
    %Cores/SMs & 6 cores, 12 threads & 48 SMs, 6144 CUDA cores & - \\
    Peak Perf. & 2.4 TFLOPS (FP32) & 19.2 TFLOPS (FP32) & 0.26 TFLOPS (FP32) \\
    On-chip Mem. & 3 MB L2, 16 MB L3 & 6 MB L2 Cache & 4.5 MB \\
    Memory BW & 50 GB/s (DDR4) & 448 GB/s (GDDR6) & 19.2 GB/s (DDR4) \\
    \bottomrule
  \end{tabular}}
\end{table}

\subsection{Evaluation Metrics}
We report four key metrics:  
(i) \emph{Classification accuracy} (\%): fraction of correctly predicted labels against ground truth;  
(ii) \emph{Latency} (ms/graph): average end-to-end inference time per graph at batch size 1, measured from the arrival of the graph’s adjacency and feature matrices to the prediction of class label;
(iii) \emph{Power} (W): average device power during steady-state inference, including static and dynamic components;  
(iv) \emph{Energy Efficiency} (mJ/graph): average energy per inference, computed as the product of measured power and latency.

\subsection{Results and Analysis}

%\subsubsection{\textbf{CPU / GPU Profiling}}
%\label{sec:cpu_gpu_breakdown}

%Figure~\ref{fig:cpu_gpu_breakdown} presents a detailed execution-time breakdown of the CPU and GPU baselines on the DD dataset. On both platforms, inference latency is dominated by \emph{memory-bound, irregular-access stages}. In particular, codebook lookup and Nyström computation account for over 50\% of total CPU execution time and over 70\% of GPU execution time. These stages involve fine-grained, data-dependent memory accesses with poor locality and high contention, which severely limit effective parallelism on general-purpose architectures. This profiling motivates our FPGA-based design, which replaces these bottlenecks with fully pipelined, banked datapaths tailored to the Nyström-HDC workload, enabling low inference latency and high energy efficiency under strict edge constraints.

%\begin{figure}
%  \centering
%  \includegraphics[width=\columnwidth]{figures/cpu_gpu_breakdown.pdf}
%  \caption{Execution-time breakdown of CPU and GPU baselines on the DD dataset. Codebook lookup and Nyström computation dominate runtime due to irregular, memory-bound access patterns, motivating our custom HyperX accelerator architecture.}
%  \label{fig:cpu_gpu_breakdown}
%\end{figure}

\subsubsection{\textbf{Latency}}
%We first evaluate the end-to-end inference latency of \textbf{HyperX} on the ZCU104 FPGA, compared against CPU and GPU baselines across the eight TU datasets. 
For FPGA, cycle-accurate latency is obtained from the integrated analyzer in the AMD Vitis Unified IDE v2024.2.
%toolchain, following HLS synthesis and implementation. 
%These estimates capture both computation and memory access overhead under the dataflow scheduling model, and therefore reflect the true end-to-end execution time of the hardware design. 
CPU and GPU latencies are measured using wall-clock timing in PyTorch with batch size set to one, consistent with real-time edge inference. Table~\ref{tab:latency-with-dpp} summarizes the results, %reports absolute latencies (ms/graph) for CPU, GPU, and FPGA with and without DPP-based landmark reduction.
%, while Fig.~\ref{fig:speedup-bars} shows corresponding speedups normalized to the CPU baseline without DPP. 
across all datasets, \textbf{HyperX} consistently delivers lower latency than CPU and GPU baselines, despite both being more resource-rich and compute-capable. On average, FPGA latency is reduced by $4.5\times$ relative to CPU and $2.2\times$ relative to GPU. The low latency is due to the FPGA-specific optimizations, such as streaming dataflow to decouple DDR from compute and BRAM-banked on-chip memory for conflict-free memory access.
%, validating the efficiency of the architecture. 
DPP further reduces per-inference latency by $25$ to $40\%$ by pruning redundant landmarks, which reduces memory transfers of $\mathbf{P}_{\mathrm{nys}}$.% and accelerates histogram updates.
%For example, on the ENZYMES dataset, FPGA latency drops from $0.61$ ms to $0.45$ ms with DPP, a $1.36\times$ reduction on top of the $7.7\times$ speedup already achieved over CPU. Similarly, on BZR, HyperX achieves $8.9\times$ speedup with DPP, the highest across all datasets. 
%Overall, \textbf{HyperX} achieves sub-millisecond inference on most datasets, with peak speedups of $8.9\times$ over CPU and $3\times$ over GPU. 
%The results demonstrate that HyperX achieves real-time performance on edge hardware while preserving a strong latency–energy efficiency advantage over CPU and GPU platforms.

% Latency (ms) with speedup in parentheses (relative to CPU w/o DPP)
% File also saved as: sandbox:/mnt/data/latency_table_snippet.tex
\begin{table}
  \centering
  \caption{End-to-end latency (ms) per graph. Speedup in parentheses is relative to the CPU baseline (without DPP).}
  \label{tab:latency-with-dpp}
  \resizebox{\columnwidth}{!}{%
  \begin{tabular}{lcccccc}
    \toprule
    \textbf{Dataset} & \textbf{CPU} & \textbf{CPU+DPP} & \textbf{GPU} & \textbf{GPU+DPP} & \textbf{FPGA} & \textbf{FPGA+DPP} \\
    \midrule
    DD & 7.47 (1.00x) & 6.11 (1.22x) & 3.00 (2.49x) & 3.00 (2.49x) & 1.80 (4.15x) & 1.65 (4.53x) \\
    ENZYMES & 4.71 (1.00x) & 2.55 (1.85x) & 1.77 (2.66x) & 1.60 (2.94x) & 0.61 (7.72x) & 0.45 (10.47x) \\
    MUTAG & 5.13 (1.00x) & 3.87 (1.33x) & 5.80 (0.88x) & 4.90 (1.05x) & 1.47 (3.49x) & 1.19 (4.31x) \\
    NCI1 & 5.04 (1.00x) & 4.23 (1.19x) & 2.70 (1.87x) & 2.60 (1.94x) & 0.98 (5.14x) & 0.61 (8.26x) \\
    BZR & 2.85 (1.00x) & 2.29 (1.24x) & 1.70 (1.67x) & 1.60 (1.78x) & 0.54 (5.27x) & 0.32 (8.89x) \\
    COX2 & 5.26 (1.00x) & 4.68 (1.12x) & 7.30 (0.72x) & 6.70 (0.79x) & 1.45 (3.63x) & 1.05 (5.01x) \\
    NCI109 & 4.26 (1.00x) & 3.44 (1.24x) & 2.50 (1.70x) & 2.60 (1.64x) & 1.07 (3.98x) & 0.69 (6.17x) \\
    Mutagenicity & 3.57 (1.00x) & 3.01 (1.19x) & 1.80 (1.98x) & 1.70 (2.10x) & 0.79 (4.52x) & 0.50 (7.13x) \\
    \bottomrule
  \end{tabular}%
}
\end{table}

\subsubsection{\textbf{Energy Efficiency}}
Table~\ref{tab:energy_efficiency} compares throughput, power, and energy efficiency across platforms. CPU power is measured with a plug-in power meter, model PM01-US, GPU power using the nvidia-smi~\cite{nvidiasmi} command line tool, and FPGA power from the AMD Vitis Unified IDE~\cite{AMD_Vitis_Unified_IDE_2024_2} post-implementation, consistent with standard FPGA power estimation~\cite{Salamat2019F5HD}. Table~\ref{tab:energy_efficiency} reports that our FPGA design delivers orders of magnitude lower energy per graph over CPU and GPU baselines, while also increasing throughput. 
%For example on DD, the FPGA achieves \(606\,\text{graphs/s}\) at \(0.81\,\text{W}\) (\(1.33\,\text{mJ/graph}\)), versus CPU: \(163\,\text{graphs/s}\) at \(25.2\,\text{W}\) (\(154\,\text{mJ/graph}\), \(116\times\) higher energy) and GPU: \(333\,\text{graphs/s}\) at \(59\,\text{W}\) (\(177\,\text{mJ/graph}\), \(133\times\) higher energy). This corresponds to \(3.7\times\) lower latency than CPU and \(1.8\times\) than GPU (throughput-equivalent). 
Across the TU datasets, the FPGA energy is \(101\times\) to \(256\times\) lower than CPU and \(133\times\) to \(451\times\) lower than GPU.
%, with throughput gains of \(3.2\times\) to \(7.2\times\) vs. CPU and \(1.8\times\) to \(6.4\times\) vs. GPU. 
These improvements arise from our architectural choices, such as DPP-based landmark reduction, bandwidth-aware streaming of $\mathbf{P}_{\mathrm{nys}}$.
%, enabling extreme energy efficiency, which is critical for edge deployment.

\begin{table}[t]
\centering
\caption{Throughput, power, and energy efficiency with DPP on TU datasets. 
Energy ratios are normalized to FPGA (=1$\times$).}
\label{tab:energy_efficiency}
\resizebox{\columnwidth}{!}{%
\begin{tabular}{llccc}
\toprule
\textbf{Dataset} & \textbf{Device} &
\textbf{\begin{tabular}{c}Throughput \\ (graphs/s)\end{tabular}} &
\textbf{\begin{tabular}{c}Power \\ (W)\end{tabular}} &
\textbf{\begin{tabular}{c}Energy Efficiency \\ (mJ/graph)\end{tabular}} \\
\midrule
            & CPU   & 163  & 25.2 & 154 (116$\times$) \\
\textbf{DD} & GPU   & 333  & 59.0 & 177 (133$\times$) \\
            & \textbf{FPGA}  & 606  & 0.81 & 1.33 (1$\times$) \\
\midrule
                 & CPU   & 392  & 24.3 & 62 (194$\times$) \\
\textbf{Enzymes} & GPU   & 625  & 61.0 & 98 (305$\times$) \\
                 & \textbf{FPGA}  & 2222 & 0.71 & 0.32 (1$\times$) \\
\midrule
               & CPU   & 259  & 25.3 & 98 (101$\times$) \\
\textbf{Mutag} & GPU   & 204  & 60.0 & 294 (303$\times$) \\
               & \textbf{FPGA}  & 840  & 0.81 & 0.97 (1$\times$) \\
\midrule
              & CPU   & 236  & 25.5 & 108 (225$\times$) \\
\textbf{NCI1} & GPU   & 385  & 61.0 & 159 (330$\times$) \\
              & \textbf{FPGA}  & 1639 & 0.79 & 0.48 (1$\times$) \\
\midrule
             & CPU   & 437  & 24.6 & 56 (256$\times$) \\
\textbf{BZR} & GPU   & 625  & 62.0 & 99 (451$\times$) \\
             & \textbf{FPGA}  & 3125 & 0.70 & 0.22 (1$\times$) \\
\midrule
              & CPU   & 214  & 24.4 & 114 (127$\times$) \\
\textbf{COX2} & GPU   & 149  & 61.0 & 409 (454$\times$) \\
              & \textbf{FPGA}  & 952  & 0.86 & 0.90 (1$\times$) \\
\midrule
                & CPU   & 291  & 24.9 & 86 (156$\times$) \\
\textbf{NCI109} & GPU   & 385  & 60.0 & 156 (284$\times$) \\
                & \textbf{FPGA}  & 1449 & 0.79 & 0.55 (1$\times$) \\
\midrule
                      & CPU   & 332  & 24.4 & 73 (184$\times$) \\
\textbf{Mutagenicity} & GPU   & 588  & 60.0 & 102 (255$\times$) \\
                      & \textbf{FPGA}  & 2000 & 0.79 & 0.40 (1$\times$) \\
\bottomrule
\end{tabular}}
\end{table}

\subsubsection{\textbf{Impact of DPP on Accuracy and Memory}}\label{results:dpp}

\begin{figure}
  \centering
  \includegraphics[width=\columnwidth]{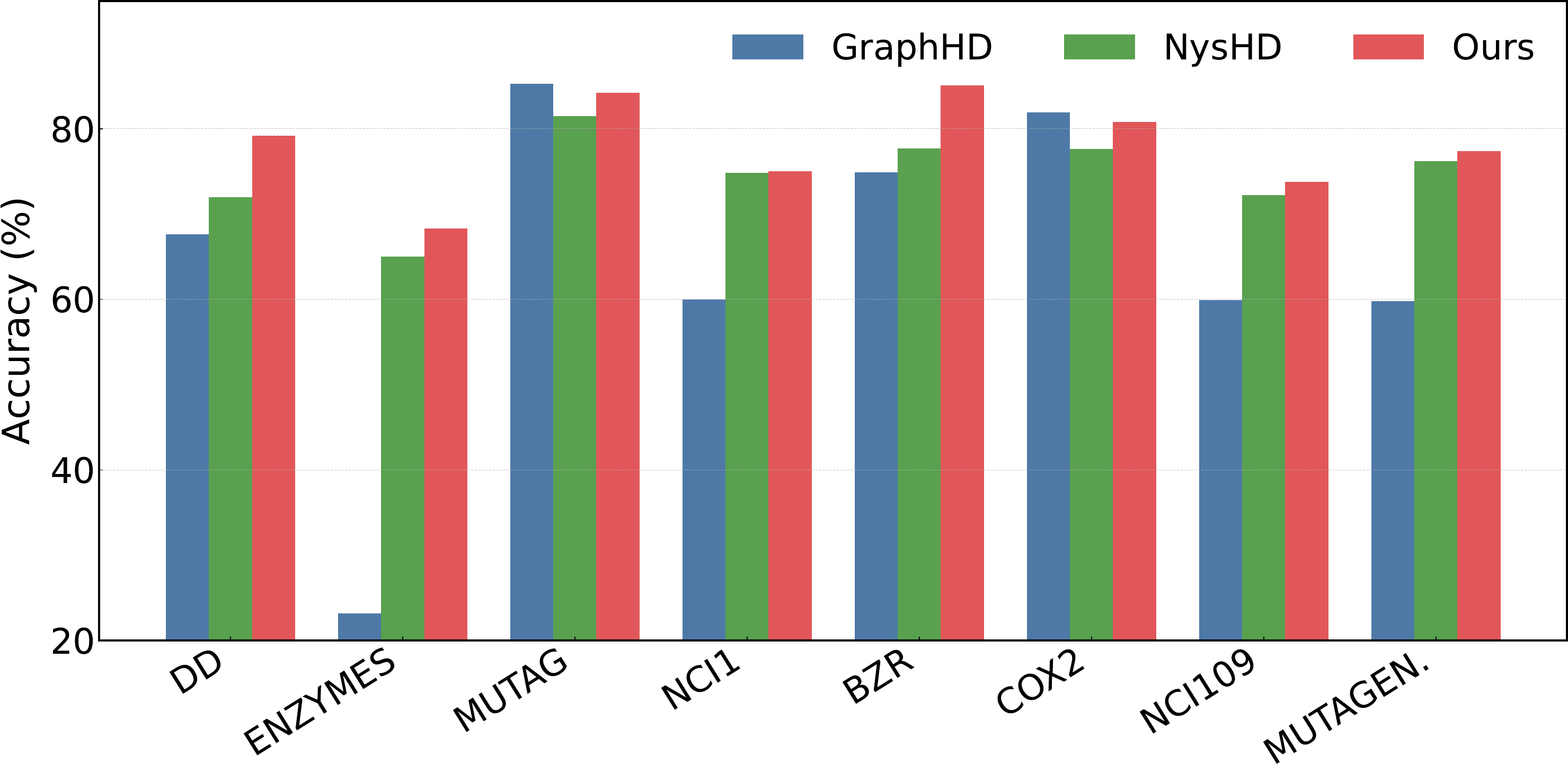}
  \caption{Classification accuracy (\%) on TU datasets.}
  \label{fig:accuracy}
\end{figure}

%We evaluate the effect of DPP-based landmark selection on both accuracy and memory in Nyström-HDC inference. 
Figure~\ref{fig:accuracy} reports classification accuracy across eight TU datasets~\cite{tudatasetcollectionbenchmarkdatasets}, comparing \textbf{HyperX} against state-of-the-art HDC baselines GraphHD~\cite{GrapHD} and NysHD~\cite{NysHD}. GraphHD results are taken from prior work~\cite{GrapHD, NysHD}, and we re-implement~\cite{NysHD} using the authors' open-source repository under identical experimental settings. Our design improves accuracy by 3.4\% on average over NysHD.
%, with the largest gains on BZR (8.0\%) and DD (7.0\%). 
We outperform both baselines on six of eight datasets; on COX2 and Mutag, GraphHD performs slightly better. This behavior reflects our hybrid uniform+DPP landmark selection. Uniform sampling introduces randomness and redundancy, which can have a significant impact on small datasets. A full DPP-based sampling could reduce this impact, but it is computationally expensive due to eigendecomposition and full kernel matrix construction. Our hybrid approach preserves most of DPP’s diversity benefits at a lower sampling cost.
%These gains arise from DPP-based landmark selection, which increases representational diversity by eliminating redundant landmark graphs.
Beyond accuracy, DPP reduces the memory required for model parameters in Algorithm~\ref{alg:end-to-end}. 
%Among them, the Nyström projection matrix $\mathbf{P}_{\mathrm{nys}}$ dominates both the storage and off-chip transfer volume. 
Table~\ref{tab:memory_dpp} shows that by eliminating redundant landmarks, DPP-based landmark reduction lowers total memory required by 37\% on average. 
%Thus, by eliminating redundant landmarks, DPP reduces the memory footprint, lowering both the storage requirements and off-chip traffic while improving accuracy.

%On edge devices, where on-chip storage is severely constrained, these parameters must otherwise be fetched from external memory for every inference, stressing the limited DDR bandwidth. 

\begin{table}
\centering
\caption{Memory consumption of model parameters.}
\resizebox{\columnwidth}{!}{
\begin{tabular}{|l|c|c|}
\hline
\textbf{Dataset} & \textbf{Memory w/o DPP (MB)} & \textbf{Memory w/ DPP (MB)} \\
\hline
DD            & 12.50 &  9.15 \,($\downarrow$ 26.8\%) \\
Enzymes       & 16.13 & 11.13 \,($\downarrow$ 31.0\%) \\
Mutag         &  7.49 &  4.62 \,($\downarrow$ 38.3\%) \\
NCI1          & 12.54 &  7.88 \,($\downarrow$ 37.2\%) \\
BZR           & 11.78 &  7.02 \,($\downarrow$ 40.4\%) \\
COX2          & 12.50 &  7.70 \,($\downarrow$ 38.4\%) \\
NCI109        & 12.50 &  6.97 \,($\downarrow$ 44.2\%) \\
Mutagenicity  & 11.86 &  7.16 \,($\downarrow$ 39.6\%) \\
\hline
\end{tabular}
}
\label{tab:memory_dpp}
\end{table}

\subsubsection{\textbf{Impact of Load Balancing (LB)}}

\begin{figure}
  \centering
  \includegraphics[width=1.0\columnwidth]{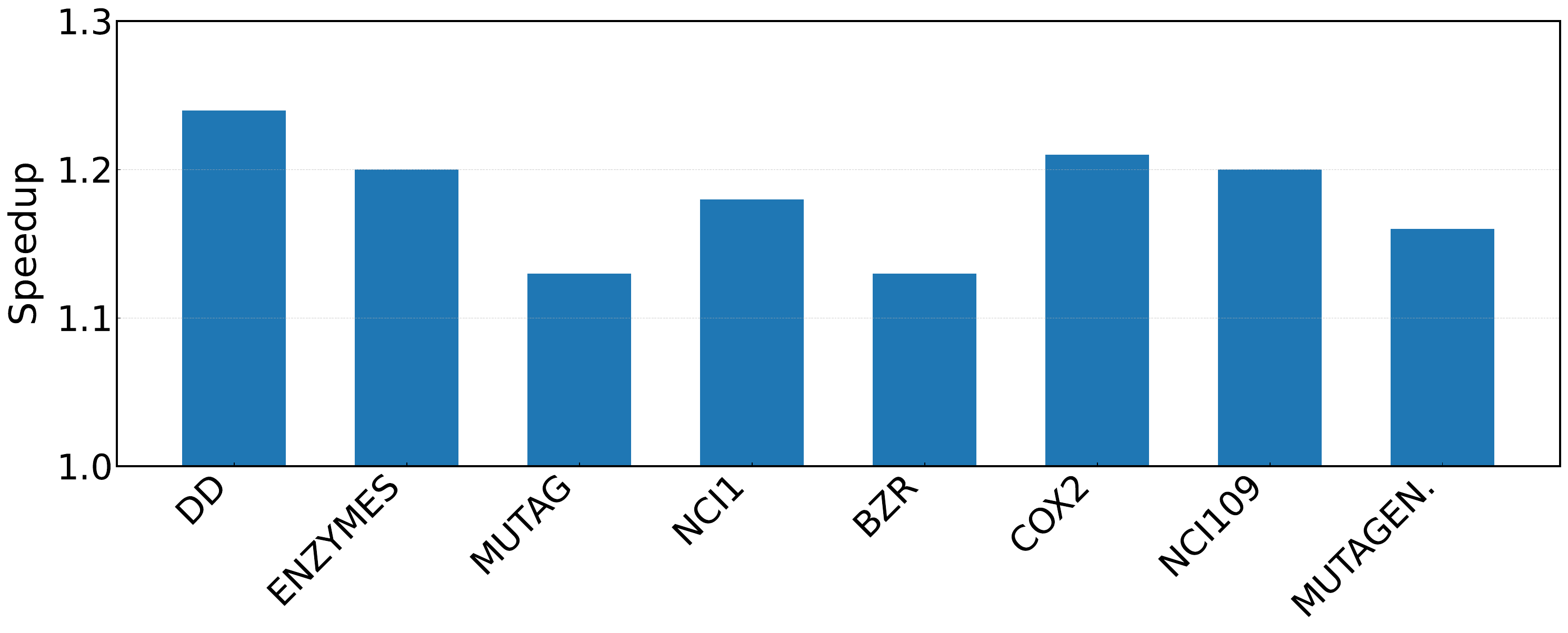}
  \caption{Effect of static load balancing in SpMV stages (LSHU/KSE): Speedup normalized to \emph{no LB} case (=1).}%; ↑ better).}
  \label{fig:lb-speedup}
\end{figure}

We quantify the benefit of the static, offline row scheduling in Section~\ref{subsec:static-lb} by comparing FPGA latency with and without load balancing. Figure~\ref{fig:lb-speedup} reports speedup normalized to the \emph{no-LB} case (=1). Across the TU datasets, we observe consistent gains of $1.19{\times}$ on average. Datasets with broader per-row nnz variance (e.g., DD, COX2) benefit the most, due to reduced PE idling and fewer BRAM-port stalls.

\subsubsection{\textbf{Ablations Studies}}
\begin{figure}
  \centering
  \includegraphics[width=0.9\columnwidth]{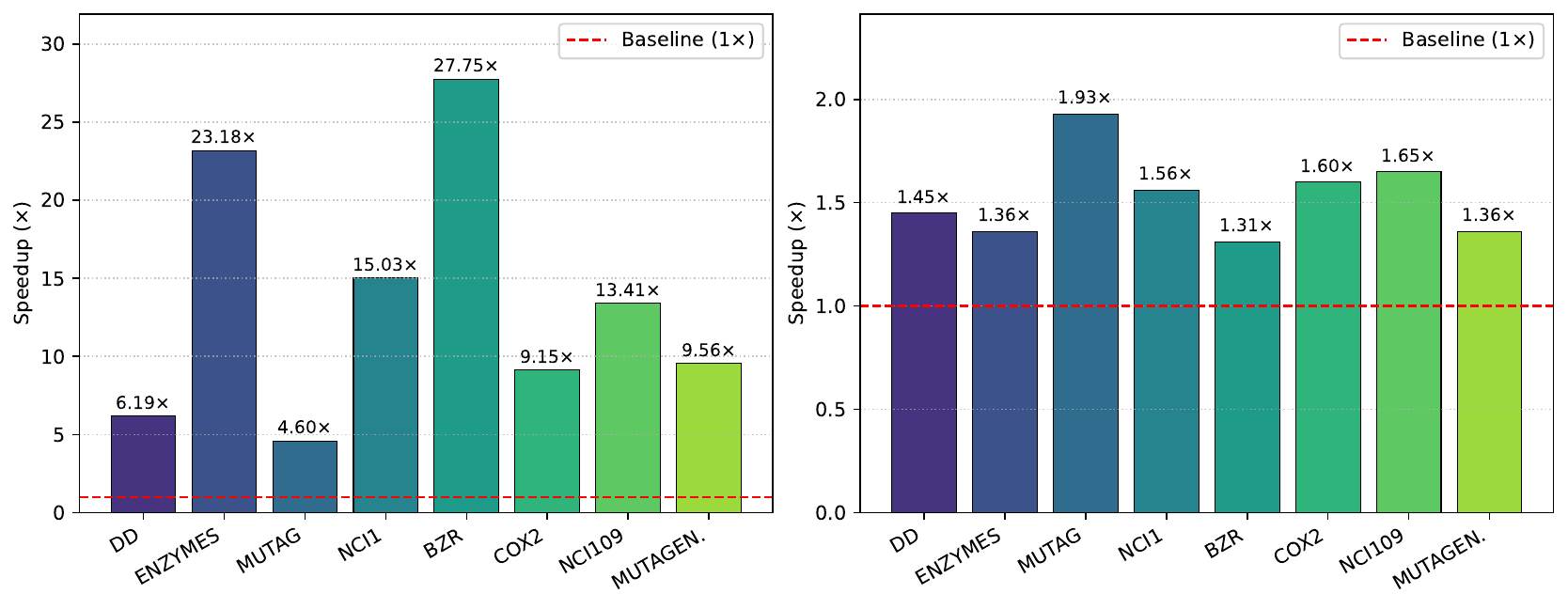}
  \caption{Ablation study showing speedup over baseline: (left) NEE ablation and (right) MPHE ablation.}
  \label{fig:ablation_speedup}
\end{figure}

Figure~\ref{fig:ablation_speedup} isolates the impact of the NEE by comparing end-to-end FPGA inference latency with and without NEE optimizations. The results show that without streaming dataflow, FIFO buffering, or 512-bit-aligned DDR access, inference latency is dominated by stalled memory accesses. Figure~\ref{fig:ablation_speedup} evaluates the impact of the MPHE by comparing it with an optimized binary-search-based codebook lookup. The results highlight our design choice of a fully pipelined $O(1)$ key-to-index mapping. MPHE introduces a small amount of auxiliary metadata to enable constant-time, conflict-free codebook lookup reported in Table~\ref{tab:mphe_memory}. Even in the largest configuration (MUTAG, 10 hops), the combined MPHE metadata and codebook size is only $20.29$\,KB ($0.16$\,Mb), which constitutes a negligible fraction of available on-chip memory.

\begin{table}
\centering
\caption{Memory overhead of MPHE metadata.}
\label{tab:mphe_memory}
\resizebox{\columnwidth}{!}{%
\begin{tabular}{lccccc}
\toprule
\textbf{Dataset} & \textbf{Hops} & \textbf{MPH Metadata} & \textbf{Codebook (KB)} & \textbf{Total (KB)} & \textbf{Overhead} \\
\midrule
DD              & 2  & 490\,B (392\,B / 98\,B)   & 4.63  & 5.11  & 10.3\% \\
Mutagenicity   & 3  & 330\,B (264\,B / 66\,B)   & 2.48  & 2.80  & 13.0\% \\
MUTAG          & 10 & 1.88\,KB (1.50\,KB / 384\,B) & 18.41 & 20.29 & 10.2\% \\
NCI109         & 5  & 840\,B (672\,B / 168\,B)  & 7.49  & 8.31  & 10.9\% \\
\bottomrule
\end{tabular}}
\end{table}
\section{Related Works}
\label{sec:related}

\noindent \textbf{Graph Classification with HDC:}  
GraphHD~\cite{GrapHD}, the first HDC approach for graph classification, encodes graph structure, but does not exploit node attributes, limiting accuracy. NysHD \cite{NysHD} improves accuracy through Nyström approximations; however, it incurs additional propagation kernel cost, relies on uniform landmark sampling leading to redundant landmarks, and does not exploit the sparsity in adjacency and histogram matrices. 

\textbf{FPGA Acceleration of HDC:} Prior hardware acceleration efforts for HDC have targeted domains such as biosignals~\cite{static-encoding-2-seizure-det}, image descriptors~\cite{static-encoding-9-hdc-framework-image-descriptors-cvpr}, and activity recognition~\cite{static-encoding-7-efficient-human-activity-recognition-hdc-imani}. E$^{3}$-HDC~\cite{E3HDC} accelerates random-projection encoding; FACH~\cite{FACH} accelerates similarity engines; DistriHD~\cite{DistriHD} scales HDC across FPGA clusters for image workloads. HyperGraf~\cite{HyperGraf} and~\cite{barkam2023reliable} apply HDC to graph reasoning tasks. These works validate hardware-accelerated HDC but target different encodings and tasks. To the best of our knowledge, \textbf{HyperX} is the first to exploit FPGA to accelerate Nyström-based HDC for graph classification. %\textbf{Nyström and Kernel Methods:}  
%The Nyström method~\cite{Williams2000Nystrom} has long been used to efficiently approximate kernel matrices, with sampling strategies ranging from uniform sampling to determinantal point processes (DPPs)~\cite{Kulesza_2012, fastdpp, Kumar2012Sampling}. \cite{NysHD} extends this line of work to HDC, showing the benefits of integrating kernel functions for feature encoding. Building on this, we employ DPP-based landmark selection with propagation-kernel similarity, and introduce an FPGA accelerator that enables Nyström-HDC graph classification with low latency and energy consumption, while significantly improving accuracy.

\section{Conclusion and Future Work}
We presented \textbf{HyperX}, the first FPGA accelerator for Nyström-based HDC graph classification, by integrating DPP-based landmark sampling, a streaming projection design, $O(1)$ MPH lookups, and statically load-balanced SpMV. Experimental results demonstrate that \textbf{HyperX} achieves significantly lower latency and higher energy efficiency compared to CPU and GPU baselines on ZCU104, while \emph{improving} classification accuracy by $3.4\%$ on average compared to prior HDC methods. Looking ahead, we plan to extend these techniques to other graph tasks such as node classification and link prediction.

\begin{acks}
This work is supported by the DEVCOM Army Research Lab (ARL) under grant W911NF-242-0194, and the National Science Foundation (NSF) under grants CSSI-2311870 and OAC-2505107.
\end{acks}

% comment out
\bibliographystyle{ACM-Reference-Format}
\bibliography{Reference}
\end{document}